%% file: main.tex
\documentclass[acmsmall]{acmart}

\input{package}
\input{macro}

\setcopyright{cc}
\setcctype{by}
\acmDOI{10.1145/3808142}
\acmYear{2026}
\acmJournal{PACMSE}
\acmVolume{3}
\acmNumber{FSE}
\acmArticle{FSE135}
\acmMonth{7}
\acmSubmissionID{fse26mainb-p469-p}
\received{2025-08-30}
\received[accepted]{2026-03-24}

\title{AdaDec: A Uncertainty-Guided Lookahead Decoding Framework for LLM-Based Code Generation}

\author{Kaifeng He}
\orcid{0009-0000-5612-3921}
\affiliation{
  \institution{Sun Yat-sen University}
  \city{Zhuhai}
  \country{China}
}
\email{hekaifeng70@gmail.com}

\author{Mingwei Liu}
\orcid{0000-0002-3462-997X}
\affiliation{
  \institution{Sun Yat-sen University}
  \city{Zhuhai}
  \country{China}
}
\email{liumw26@mail.sysu.edu.cn}

\author{Chong Wang}
\orcid{0000-0003-1424-6290}
\affiliation{
  \institution{Nanyang Technological University}
  \city{Singapore}
  \country{Singapore}
}
\email{chong.wang@ntu.edu.sg}

\author{Zike Li}
\orcid{0009-0000-0807-1805}
\affiliation{
  \institution{Sun Yat-sen University}
  \city{Zhuhai}
  \country{China}
}
\email{lizk8@mail2.sysu.edu.cn}

\author{Yanlin Wang}
\orcid{0000-0001-7761-7269}
\affiliation{
  \institution{Sun Yat-sen University}
  \city{Zhuhai}
  \country{China}
}
\email{yanlin-wang@outlook.com}

\author{Xin Peng}
\orcid{0000-0003-3376-2581}
\affiliation{
  \institution{Fudan University}
  \city{Shanghai}
  \country{China}
}
\email{pengxin@fudan.edu.cn}

\author{Zibin Zheng}
\orcid{0000-0002-7878-4330}
\affiliation{
  \institution{Sun Yat-sen University}
  \city{Zhuhai}
  \country{China}
}
\email{zhzibin@mail.sysu.edu.cn}

\authorsaddresses{
  $^\star$Corresponding authors: Mingwei Liu and Chong Wang. \\
  Kaifeng He, Mingwei Liu, Zike Li, Yanlin Wang, and Zibin Zheng are with the School of Software Engineering and the Zhuhai Key Laboratory of Trusted Large Language Models, Sun Yat-sen University, Zhuhai, China.
  Email: hekaifeng70@gmail.com, liumw26@mail.sysu.edu.cn, lizk8@mail2.sysu.edu.cn, yanlin-wang@outlook.com, zhzibin@mail.sysu.edu.cn. \\
  Chong Wang is with Nanyang Technological University, Singapore.
  Email: chong.wang@ntu.edu.sg. \\
  Xin Peng is with Fudan University, Shanghai, China.
  Email: pengxin@fudan.edu.cn.
}

\begin{document}

\input{sections/abstract}



\maketitle

\input{sections/introduction}
\input{sections/empirical}
\input{sections/approach}
\input{sections/evaluation}
\input{sections/discussion}
\input{sections/threats}
\input{sections/related_work}
\input{sections/conclusion}
\input{sections/data_availability}
\input{sections/acknowledgments}

\normalem
\bibliographystyle{ACM-Reference-Format}
\balance
\bibliography{refs}

\end{document}

%% file: package.tex
\usepackage{algorithmic}
\usepackage{graphicx}
\usepackage{textcomp}
\usepackage{xcolor}

\usepackage{makecell}
\usepackage{tabularx}
\usepackage[T1]{fontenc}
\usepackage{comment}
\usepackage{xcolor}
\usepackage{multirow}
\usepackage{array}
\usepackage{soul}
\usepackage{threeparttable}
\usepackage{balance}
\usepackage{diagbox}
\usepackage{ulem}
\usepackage{subfigure}
\usepackage{microtype} 
\usepackage{calc}

\usepackage{enumitem}
\usepackage{url}
\usepackage{xspace}

\usepackage{float}
\usepackage{stfloats}
\usepackage{adjustbox}

\usepackage{amsmath,amssymb,amsfonts}
\usepackage{algorithmic}
\usepackage{graphicx}
\usepackage{textcomp}
\usepackage{xcolor}
\usepackage{multirow}

\usepackage[most]{tcolorbox}
\usepackage{makecell}
\usepackage{threeparttable}
\usepackage{colortbl}
\usepackage{subfigure}
\usepackage{hhline}
\usepackage{pifont}

\usepackage{enumitem}

\usepackage{booktabs}
\usepackage{siunitx}
\usepackage{booktabs}

%% file: macro.tex
\usepackage[utf8]{inputenc}
\usepackage[english]{babel}
\usepackage{enumitem}
\usepackage{bm}

\newcommand\app{\textsc{AdaDec}\xspace}

\newcommand{\dsOne}{DS-1.3B\xspace}
\newcommand{\dsSix}{DS-6.7B\xspace}
\newcommand{\stThree}{ST-3B\xspace}
\newcommand{\qwThreeZ}{QW3-0.6B\xspace}
\newcommand{\qwThreeOne}{QW3-1.7B\xspace}
\newcommand{\qwThreeFour}{QW3-4B\xspace}
\newcommand{\qwThreeEight}{QW3-8B\xspace}
\newcommand{\qwCodeOne}{QW2.5-1.5B\xspace}
\newcommand{\qwCodeSeven}{QW2.5-7B\xspace}

\newcommand{\Comment}[1]{}

\definecolor{ggray}{HTML}{eff0f0}
\definecolor{gggray}{HTML}{E8E8E8}
\definecolor{ggggray}{HTML}{BEBEBE}

\newcounter{finding}
\newcommand{\finding}[1]{
  \refstepcounter{finding}
  \begin{tcolorbox}[
    enhanced,
    frame hidden,
    boxrule=0pt,
    borderline west={3pt}{0pt}{gray},
    colback=gray!15,
    left=2pt,
    right=0pt,
    top=2pt,
    bottom=2pt,
    before skip=1mm,
    after skip=1mm
  ]
    \textbf{Finding \arabic{finding}:} #1
  \end{tcolorbox}
}

\newcommand{\summary}[1]{
  \begin{tcolorbox}[
    enhanced,
    frame hidden,
    boxrule=0pt,
    borderline west={3pt}{0pt}{gray},
    colback=gray!15,
    left=2pt,
    right=0pt,
    top=2pt,
    bottom=2pt,
    before skip=1mm,
    after skip=1mm
  ]
    \textbf{Summary:} #1
  \end{tcolorbox}
}

\newcommand{\distance}{5pt}

%% file: sections/abstract.tex
\begin{abstract}
Code generation with large language models (LLMs) is highly sensitive to token selection during decoding, particularly at uncertain decision points that influence program logic. While standard strategies such as greedy decoding treat all tokens uniformly, they overlook code-specific uncertainty patterns, leading to suboptimal performance. This paper presents an empirical study revealing that many generation errors stem from token ranking mistakes at high-uncertainty steps, where the correct token is present but not top-ranked. 

Motivated by these findings, we propose \app, a lookahead-based uncertainty-guided adaptive decoding framework that integrates a token-level \textit{pause-then-rerank} mechanism driven by token uncertainty. \app learns model-specific uncertainty thresholds and applies a lookahead-based reranking strategy when uncertainty is high. 
Experiments on HumanEval+, MBPP+, and DevEval benchmarks show that \app improves Pass@1 accuracy by up to 20.9\% in absolute terms over greedy decoding. More importantly, it consistently outperforms both competitive baselines like Beam Search and state-of-the-art adaptive decoding methods such as AdapT, while maintaining high efficiency through selective, uncertainty-triggered pausing. 
Our results highlight the promise of uncertainty-aware adaptive decoding for improving both the reliability and efficiency of LLM-based code generation.
\end{abstract}

%% file: sections/introduction.tex
\section{Introduction}\label{sec:intro}

Large language models (LLMs) have significantly advanced the field of code generation by learning expressive representations of programming syntax and semantics from large-scale corpora of source code and documentation~\cite{LiJia2023SKCODER,di2025enhancingcodegenerationbidirectional,tian2024fixinglargelanguagemodels,jiang2025rocodeintegratingbacktrackingmechanism,lin2024soen101codegenerationemulating, cao2026rigorreliabilityreproducibilitymatter}. Models such as CodeLlama~\cite{rozière2024codellamaopenfoundation} and DeepSeek-Coder~\cite{deepseek-coder}—pre-trained on billions of lines of code—exhibit strong capabilities in producing syntactically correct and context-aware code snippets from natural language prompts~\cite{yuan2023evaluating, sun2024enhancing, ugare2024improving, li2025preliminarystudyrobustnesscode}. These capabilities have catalyzed a range of intelligent software engineering tasks, including code completion, automated bug fixing~\cite{fu_vulrepair_2023, jin_inferfix_2023, xia_less_2022}, and test-case generation, ultimately reducing developer effort and accelerating the software development process~\cite{jiang2024surveylargelanguagemodels,huynh2025largelanguagemodelscode,  liu2026shortcoderknowledgeaugmentedsyntaxoptimization}.

At the core of code generation with large language models (LLMs) lies the decoding process, which translates learned latent representations into discrete, executable token sequences~\cite{Humaneval}. Decoding for code generation, however, faces domain-specific challenges. Code exhibits distinctive patterns of token uncertainty: certain tokens (e.g., identifiers or control keywords at the start of a line) are intrinsically ambiguous because their correct choice depends on \textbf{future logic}, while others (e.g., syntactically constrained keywords or repeated identifiers) are relatively deterministic. Crucially, mistakes at a few high-uncertainty points can yield irreversible semantic errors (for example, selecting ``for'' versus ``return'' at line start fundamentally alters control flow). Figure~\ref{fig:intro_example} provides an example. Therefore, the choice of decoding strategy—how the next token is selected or processed—plays a decisive role in balancing efficiency, diversity, and reliability.  

\input{figures/intro_example}

A range of strategies have been proposed: deterministic greedy decoding~\cite{Radford2019LanguageMA} is simple and efficient but prone to premature commitment to suboptimal tokens; 
beam search~\cite{beamsearch} improves results by maintaining multiple candidates but incurs high computational overhead; 
sampling-based approaches such as top-\(k\)~\cite{DBLP:conf/iclr/nsample} and nucleus (top-\(p\))~\cite{DBLP:conf/iclr/nsample} sampling enhance diversity but often introduce syntactic or semantic errors that are unacceptable for code; and uncertainty-based methods aim to exploit token-level difficulty signals. For instance, AdapT~\cite{AdapT} estimates uncertainty by computing cross-entropy loss between model predictions and ground truth tokens, finding that tokens with higher loss values tend to cluster at line starts. Based on this \textit{pre-identified} pattern rather than dynamic signals during decoding, AdapT assigns higher temperatures to line-initial tokens, but its reliance on randomness makes gains unstable and limited. By contrast, Uncert-CoT~\cite{UnCert-CoT} dynamically measures uncertainty at decoding time, employing Shannon entropy~\cite{Shannon} or probability difference (the gap between the top-1 and top-2 token probabilities) to decide whether a line-initial token is highly uncertain, in which case single-line chain-of-thought reasoning is triggered. While this avoids uniform treatment of all line starts, it overlooks uncertain tokens beyond line boundaries and has shown only marginal improvements in practice. Overall, existing methods either treat all decoding steps uniformly or rely on coarse heuristics tied to line structure, and current uncertainty-based decoding approaches have not yet explored which uncertainty signals are most appropriate to guide intervention at decoding time.

To address this gap, we conduct an empirical study to better understand token-level uncertainty during the decoding process in LLM-based code generation. Using DeepSeek-Coder~\cite{deepseek-coder} and the HumanEval dataset~\cite{Humaneval}, we analyze decoding failures by comparing model predictions that fail test cases with their corresponding ground-truth solutions. \textbf{Finding 1}: The analysis reveals that many errors stem from local ranking mistakes—where the correct token exists in the candidate set but is not ranked first—thereby causing logic drift. To further investigate this phenomenon, we perform next-token prediction on 164 HumanEval problems, recording decoding information and the rank of the ground-truth token at each step. This enables us to statistically identify the locations of logic drift points and analyze the relationship between logic drift and entropy. \textbf{Finding 2}: The results suggest that entropy can serve as a useful signal for detecting logic drift and guiding adaptive pausing and reranking. However, due to variability across models, a fixed entropy threshold is suboptimal, highlighting the need for a dynamic, model-aware thresholding mechanism.

Based on these findings, we introduce \app, an uncertainty-guided adaptive decoding framework for code generation with large language models. \app enhances standard decoding (e.g., greedy decoding, sampling) by incorporating a \textbf{pause-then-rerank} mechanism that reacts to model uncertainty, measured by Shannon entropy at each step. When uncertainty exceeds a threshold, decoding is paused and a reranking procedure is triggered. Because a fixed threshold does not generalize well across different models, \app learns this threshold through a data-driven logistic regression approach. During reranking, \app selects the top candidate tokens and evaluates them using a lookahead strategy inspired by A* search heuristics and a prior constrained generation method~\cite{lu2021neurologic}, scoring each candidate based on expected future continuations. The highest-scoring token is chosen for generation, and the process repeats until the sequence is complete. In contrast to prior approaches that either treat all decoding steps uniformly or rely on coarse heuristics tied to line structure, \app systematically explores Shannon entropy as a token-level uncertainty signal and applies dynamic, token-level intervention across both line-initial and in-line tokens only at high-uncertainty steps, addressing the identified gap and enabling targeted correction of the most consequential decision points.

We conduct extensive experiments on the HumanEval+, MBPP+ and DevEval benchmarks using models ranging from 0.6B to 8B parameters. Our uncertainty-guided adaptive decoding framework, \app, consistently improves Pass\@1 accuracy by up to 20.9\% over Greedy decoding and consistently outperforms state-of-the-art adaptive decoding methods such as AdapT. In practice, \app selectively pauses decoding at a low rate (averaging 6.75\%), and the targeted lookahead introduces moderate additional latency, effectively balancing quality and efficiency. Ablation studies confirm that the learned entropy threshold accurately identifies uncertain decoding steps, and empirical comparisons show that using the learned threshold outperforms a fixed threshold, particularly for models where the fixed threshold deviates substantially from the learned value. Overall, \app generalizes well across model scales and datasets, significantly improving code generation accuracy without incurring excessive runtime overhead.

In summary, this paper makes the following contributions:
\begin{itemize}
    \item An empirical study that reveals many code generation errors arise from local token ranking mistakes, not only occurring at line beginnings but also sometimes within lines, and that Shannon entropy serves as a useful signal for detecting uncertain decoding steps prone to logic drift.
    \item An adaptive decoding framework, \app, which employs an entropy-guided pause-then-rerank mechanism with learned, model-specific thresholds and a lookahead strategy to improve generation quality.
    \item Experimental results that show \app improves Pass@1 accuracy by up to 20.9\% over Greedy decoding, achieves high efficiency by selectively pausing decoding at a low rate, and generalizes effectively across diverse model sizes and benchmark datasets.
\end{itemize}

%% file: figures/intro_example.tex
\begin{figure}[htbp]
    \centering
    \includegraphics[width=0.8\columnwidth]{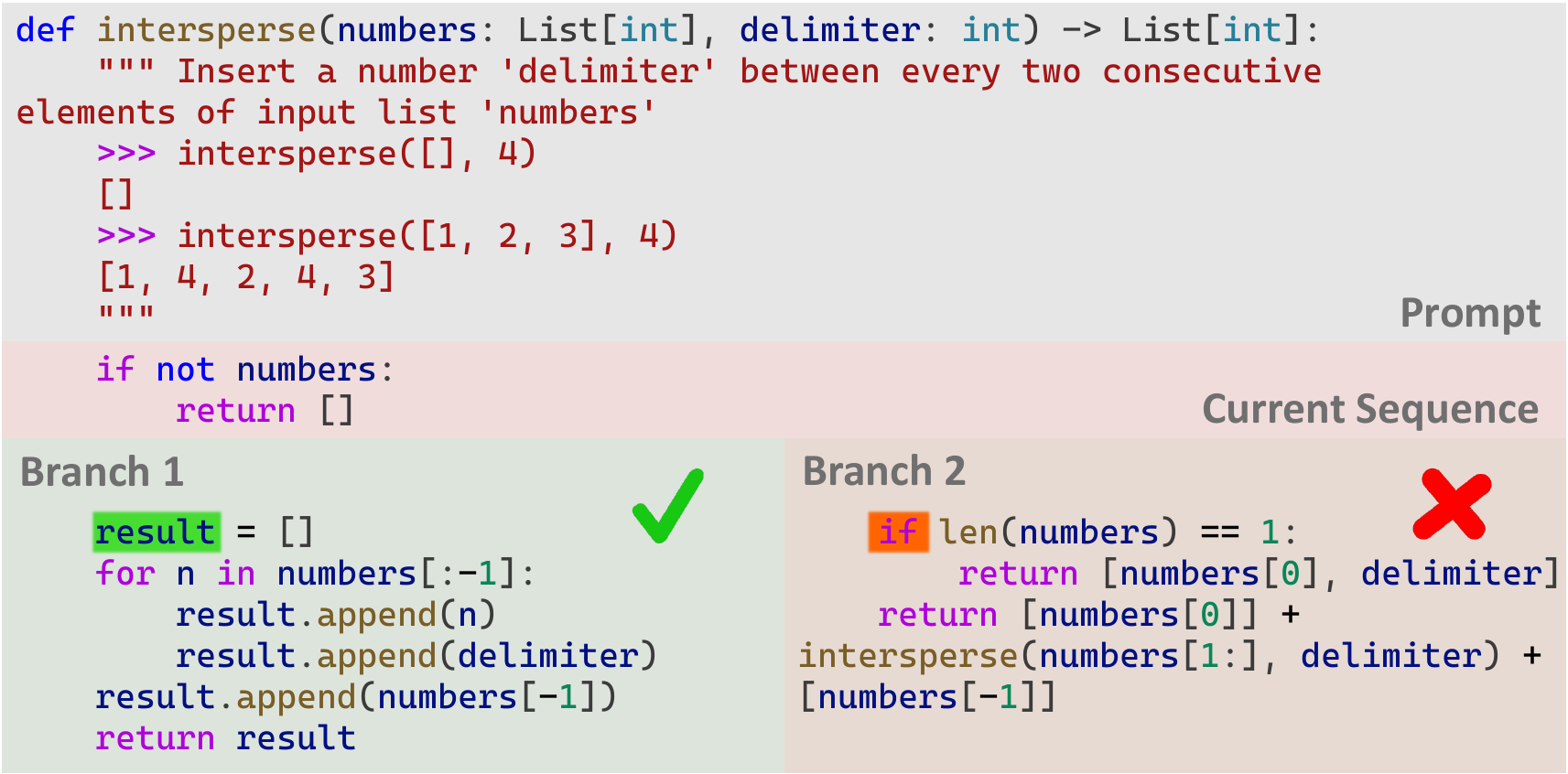}
    \Description{}
    \caption{Illustration of irreversible semantic errors in code due to high-uncertainty points.}
    \label{fig:intro_example}
\end{figure}

%% file: sections/empirical.tex
\section{Empirical Study}\label{sec:empirical}

We conduct an empirical study to investigate the characteristics of token uncertainty in LLM-based code generation and analyze the feasibility of integrating token uncertainty into decoding process. Specifically, the study aims to answer the following research questions:

\begin{itemize}
    \item \textbf{RQ1:} Where do logic drifts occur in code, and how are they influenced by token uncertainty?
    \item \textbf{RQ2:} How can token uncertainty be leveraged to improve token selection during decoding?
\end{itemize}

\subsection{Study Setup}

\subsubsection{Models}
To ensure the generality of our findings, we conduct our study on a diverse set of open-source models ranging from 0.6B to 8B parameters, accessed via Hugging Face. These encompass various architectures and instruction-tuning schemes. Specifically, we include code-specialized models: DeepSeek-Coder-1.3B/6.7B-Instruct (\dsOne, \dsSix)~\cite{deepseek-coder}, StableCode-Instruct-3B (\stThree)~\cite{stable-code-instruct-3b}, and Qwen2.5-Coder-1.5B/7B-Instruct (\qwCodeOne, \qwCodeSeven)~\cite{qwen25coder}. We also include general-purpose models featuring dynamic reasoning capabilities: the Qwen3 family, including the 0.6B (\qwThreeZ), 1.7B (\qwThreeOne), 4B (\qwThreeFour), and 8B (\qwThreeEight) variants~\cite{qwen3technicalreport}.

\subsubsection{Implementation}
Our implementation is built with PyTorch and the Hugging Face Transformers library. Decoding is performed without batching (i.e., using a batch size of 1) to ensure independent processing of each instance. The maximum generation length is capped at 1,024 tokens for all methods to enable fair comparison across models.

Experiments are conducted on a machine equipped with NVIDIA RTX 5880 Ada Generation GPUs (48 GB VRAM each) and an NVIDIA A100 GPU (80 GB VRAM), supported by NVIDIA Driver 550.163.01 and CUDA 12.4.

\subsection{RQ1: Where do logic drifts occur in code, and how are they influenced by token uncertainty?}
To address this research question, we first use \dsSix as a representative model to generate code for 164 programming problems in the widely used HumanEval dataset~\cite{Humaneval}. We begin by analyzing the locations of \textit{logic drifts}—points during decoding where the generated sequence diverges from the correct solution—which often lead to incorrect program behavior. We then investigate how token-level uncertainty, quantified using Shannon entropy~\cite{Shannon}, relates to the occurrence of such drifts.

\subsubsection{Method}
For each programming problem in the HumanEval dataset, we use \dsSix to generate code snippets using a greedy decoding strategy. 
We then filter out the generations that successfully pass all test cases. Based on an initial human analysis, we observe that in several cases, the model begins with a correct sequence but introduces a critical error—often in the form of incorrect code structures—at a specific step. For example, instead of generating a \texttt{for} loop to iterate over remaining elements, the model might incorrectly rank the \texttt{return} token at the top position, prematurely terminating the control flow. At these error points, the correct next token is not missing from the candidate set but is ranked below the top-1 position (commonly second or third), resulting in an irreversible deviation from the correct logic. This suggests that some generation failures are caused by local ranking errors at specific decoding steps, ultimately leading to incorrect code behavior. We refer to such steps as logic \textit{drift points} during decoding and aim to identify them.

Given the high cost of fully manual annotation, we develop a semi-automatic approach that combines abstract syntax tree (AST) comparison with human verification and fine-grained token-level annotation. To ensure annotation focuses only on faulty generations, we first filter out all generated samples that can pass the tests. Subsequently, we construct ASTs for both the ground-truth and generated code, and normalize them by removing identifiers and comments. This normalization ensures that the comparison focuses solely on structural differences. We then perform a parallel traversal to compare the two normalized ASTs node by node. For instance, a structural discrepancy is flagged when:
\begin{enumerate}
    \item[(1)] control-flow node types differ (e.g., \texttt{if} vs. \texttt{for});
    \item[(2)] relational or boolean operators differ (e.g., \texttt{>} vs. \texttt{>=});
    \item[(3)] loop or branching boundaries diverge;
    \item[(4)] indentation-aligned block structures differ.
\end{enumerate}
Any identified mismatch is traced back to its corresponding line in the original code and extracted as a candidate drift point.

For each candidate line, we further determine whether the drift occurs at the beginning of the line and record this for statistical analysis. Finally, each candidate is reviewed by a human annotator to confirm the validity of the drift and to annotate the precise start token. To ensure consistency, annotators adhered to the following explicit guidelines:
\begin{enumerate}
    \item[(i)] focus strictly on structural (AST-level) differences;
    \item[(ii)] ignore lexical or superficial variations (e.g., variable names, comments);
    \item[(iii)] label the earliest divergence point;
    \item[(iv)] treat alternative valid implementations as non-drift unless the semantics fundamentally change.
\end{enumerate}
For instance, a mismatch between the generated line \texttt{if shift > len(s):} and the ground-truth line \texttt{if shift >= len(digits):} triggers a drift point at the position of the operator \texttt{>=}.

In addition to the identified drift points, we collect all other non-drift decoding steps from the same generated code snippets. We compute the Shannon entropy for each step and compare the entropy distribution of drift points with that of non-drift steps. This comparison allows us to investigate the relationship between logic drift and token uncertainty, as measured by Shannon entropy. Given a probability distribution $\bm{p}$, the corresponding entropy $H(\bm{p})$ is calculated as follows:
\[
H(\bm{p}) = -\sum_{p_i \in \bm{p}} p_i \log p_i
\]

\subsubsection{Results}
We identify a total of 46 drift points and 2,889 non-drift decoding steps across 46 HumanEval problems. Statistical analysis shows that divergences at line start account for 86.96\% of all drift points, while 13.04\% occur within lines. This observation suggests that focusing solely on line-initial tokens for uncertainty-based intervention is insufficient, as it ignores more than 10\% of logic drift points. Figure~\ref{fig:entropy_analysis} presents a boxplot comparing the entropy distribution at drift points with that of all non-drift steps. For clarity, outliers are omitted in boxplots. The entropy values at drift points are markedly higher: the median entropy is 1.26, with a mean of 1.39, compared to a much lower median of 0.03 and mean of 0.30 for the rest. The interquartile range (IQR) for drift points is also shifted higher (0.63–2.01) relative to that of non-drift steps (0.003–0.34). The results show that tokens at drift points exhibit significantly higher entropy values compared to the rest, suggesting a clear relationship between entropy and logic drift.

\input{figures/entropy_analysis}

\finding{This empirical pattern indicates that focusing solely on line-start tokens for uncertainty-based intervention is insufficient, motivating the search for a suitable uncertainty signal that monitors all tokens and triggers targeted intervention at high-uncertainty steps. Subsequent analysis shows that token uncertainty measured by Shannon entropy is strongly correlated with logic drift, providing a useful signal for identifying potential drift points and predicting code behavior deviations during decoding.}

\subsection{RQ2: How can token uncertainty be leveraged to improve token selection during decoding?}
To integrate token uncertainty into the token selection process and enhance the likelihood of generating correct code, we further investigate the relationship between entropy and the rank positions of expected tokens during decoding.

\subsubsection{Method}
We perform next-token prediction using 164 problems and their corresponding ground-truth solutions from the HumanEval dataset. Specifically, given a problem $P$ and its ground-truth solution $S$, we tokenize them into sequences $\mathrm{T}^P$ and $\mathrm{T}^S$, respectively. At each decoding step $k$, we feed the concatenated sequence $\mathrm{T}^P \oplus \mathrm{T}^S_{:k}$ into the LLM to obtain a probability distribution over the next token. Here, $\mathrm{T}^S_{:k}$ denotes the prefix of the ground-truth solution containing the first $k$ tokens, and $\oplus$ indicates sequence concatenation. By iterating over all positions $k$ from 1 to the length of $\mathrm{T}^S$, we collect a set of predicted probability distributions. For each prediction, we compute the Shannon entropy of the distribution and record the rank of the expected (ground-truth) token within it.

We apply this next-token prediction procedure to a range of open-source LLMs, including five code-specific models, \dsOne, \dsSix, \stThree, \qwCodeOne, and \qwCodeSeven, as well as four general-purpose models: \qwThreeZ, \qwThreeOne, \qwThreeFour, and \qwThreeEight.
For Qwen3, we disable the reasoning-specific settings (e.g., thinking mode).
In total, we collect 180,879 token-level probability distributions for two DeepSeek-Coder models, 169,496 for StableCode, and 129,165 for Qwen series.

\subsubsection{Results}
We compute the Spearman correlation between entropy and the rank of the ground-truth token across all models. As shown in Table~\ref{tab:entropy_spearman}, the results consistently exhibit a positive correlation, further reinforcing the notion that entropy serves as a meaningful signal for identifying uncertain decoding steps. We omit p-values from the table, as the large sample sizes (approximately 130,000 to 180,000 tokens per model) yield statistically significant results with p-values effectively indistinguishable from zero.

\input{tables/entropy_spearman}

\begin{figure}[t]
\centering
\begin{minipage}{0.49\columnwidth}
  \centering
  \includegraphics[width=\linewidth]{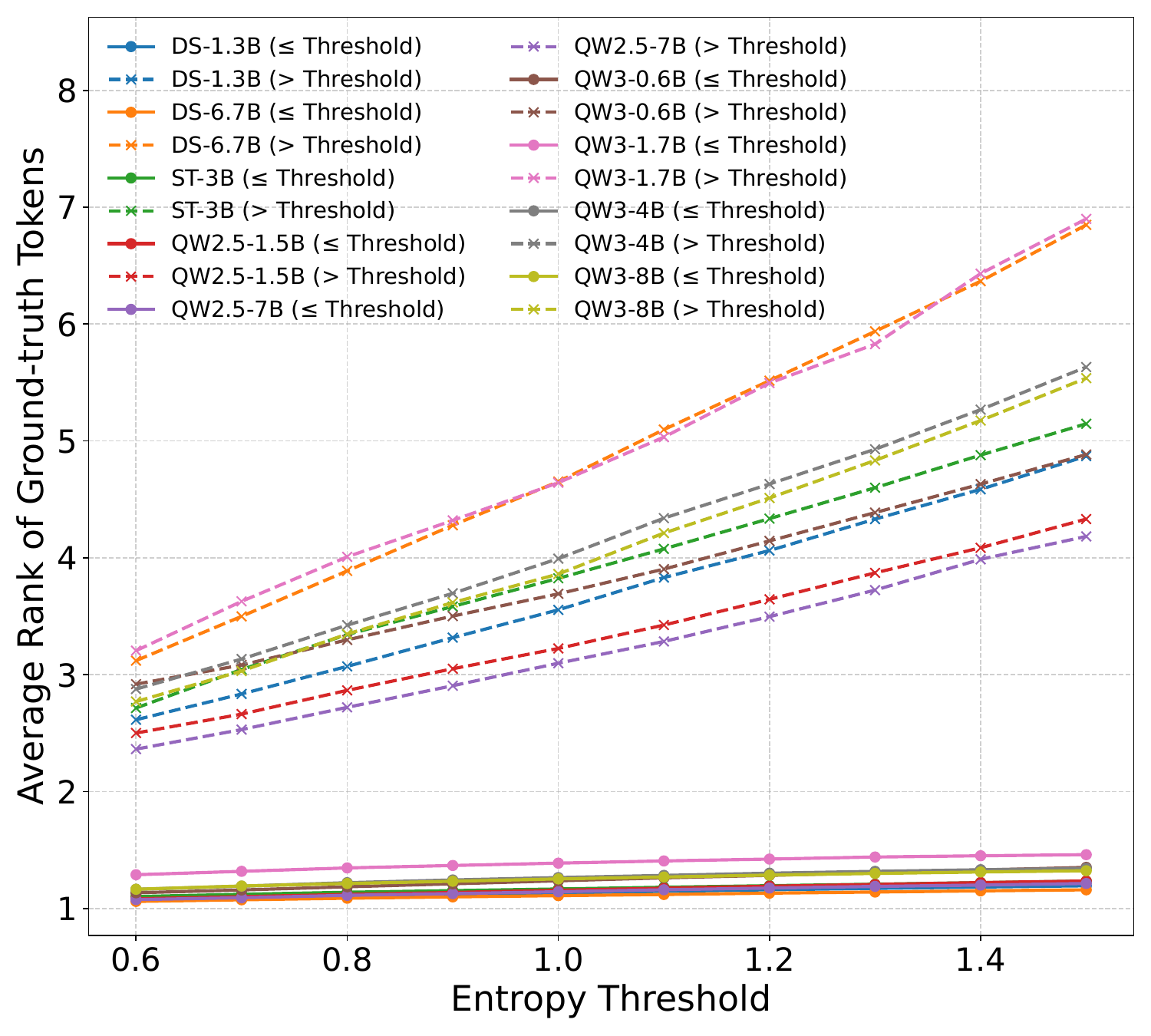}
  \Description{}
  \caption{Change in the average rank of ground-truth tokens above and below a given entropy threshold, as the threshold increases.}
  \label{fig:thres_avg_rank}
\end{minipage}\hfill
\begin{minipage}{0.49\columnwidth}
  \centering
  \includegraphics[width=\linewidth]{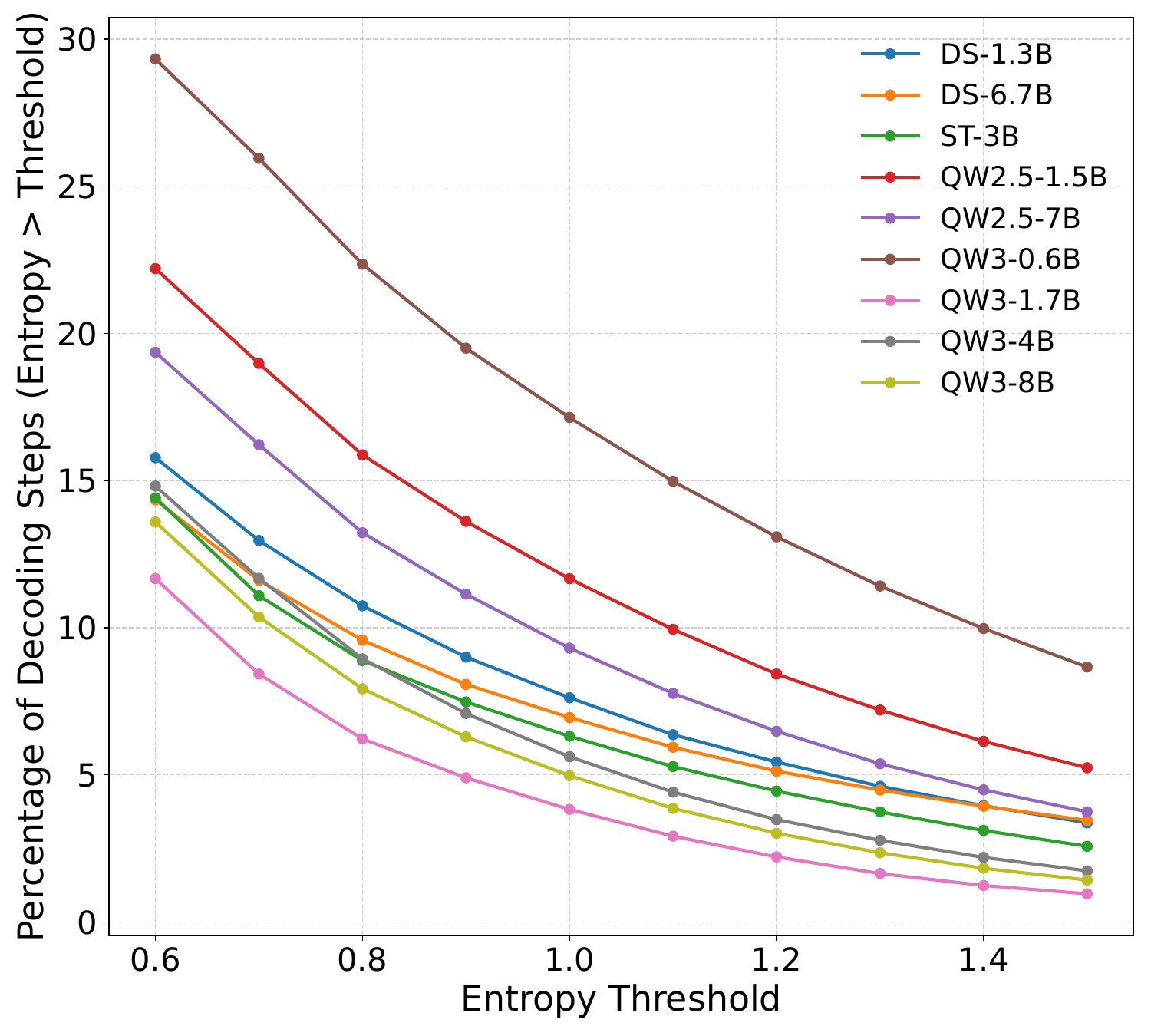}
  \Description{}
  \caption{Change in the percentage of decoding steps exceeding a given entropy threshold as the threshold increases.}
  \label{fig:thres_tok_pct}
\end{minipage}
\end{figure}

Figure~\ref{fig:thres_tok_pct} shows how the percentage of decoding steps with entropy exceeding a given threshold changes as the threshold increases. Figure~\ref{fig:thres_avg_rank} illustrates the corresponding change in the average rank of ground-truth tokens, separated by whether the entropy at each step is above or below the threshold. These plots reveal that different models exhibit varying sensitivity to entropy thresholds. For example, at a threshold of 1.2, \qwThreeZ has approximately 13\% of decoding steps exceeding the threshold, while other models remain below 10\%. In terms of average rank, two models (\dsSix and \qwThreeOne) exceed a value of 5.5, and the rest fall in the range of 3.5 to 4.7.

\finding{The observed correlation between entropy and the rank of the ground-truth token suggests that entropy can be used as an indicator to adaptively \textbf{pause} the decoding process and \textbf{rerank} uncertain tokens. However, our entropy percentile analysis shows that it is difficult to define a universal, fixed entropy threshold across all models that effectively balances pause frequency and the number of reranking candidates. This highlights the need for a \textbf{dynamic} mechanism to determine the model-specific entropy threshold.}

%% file: figures/entropy_analysis.tex
\begin{figure}[htbp]
    \centering
    \includegraphics[width=0.8\columnwidth]{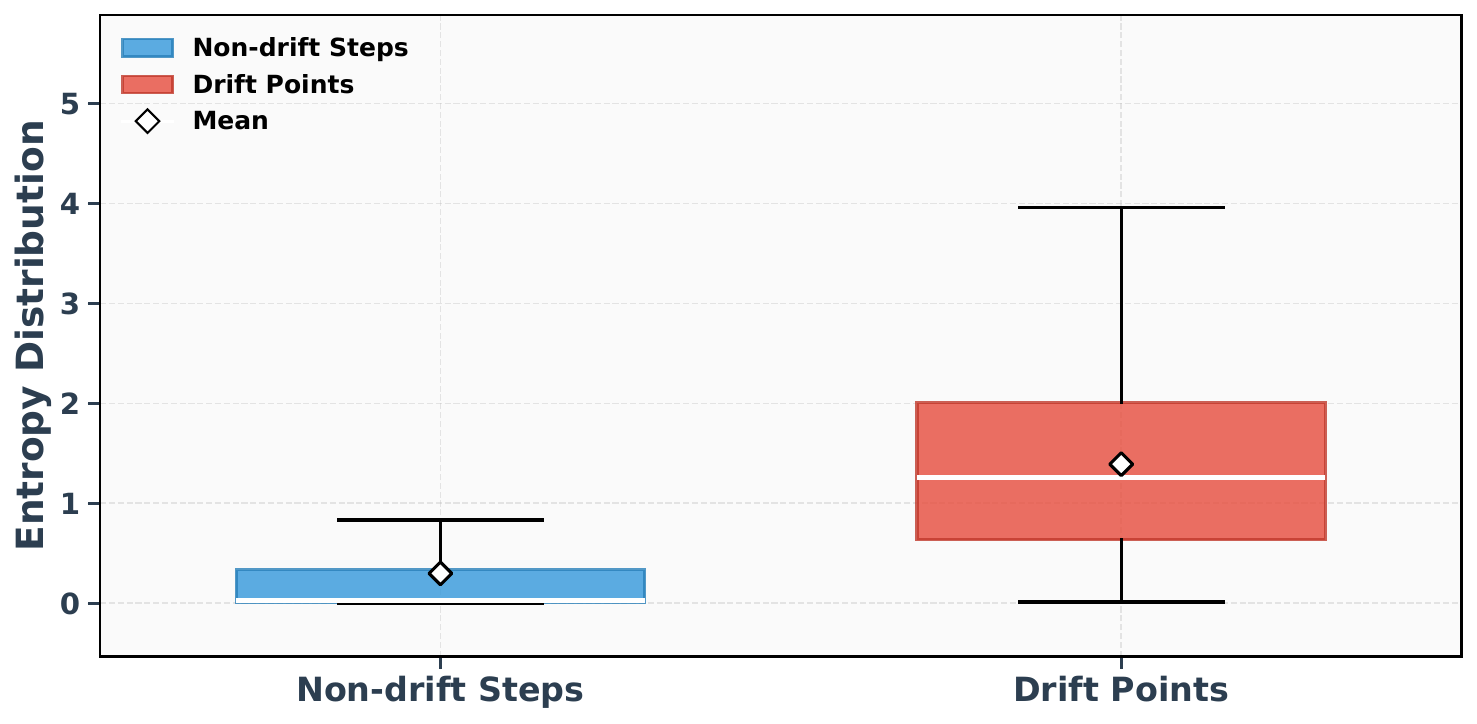}
    \Description{}
    \caption{Entropy comparison between drift points and non-drift decoding steps.}
    \label{fig:entropy_analysis}
\end{figure}

%% file: tables/entropy_spearman.tex
\begin{table}[ht]
\centering
\renewcommand{\arraystretch}{1.2}
\caption{Spearman correlation $\bm{\rho}$ between entropy and the rank of ground-truth tokens.}
\label{tab:entropy_spearman}
\resizebox{\columnwidth}{!}{
\begin{tabular}{lccccccccc}
\toprule
\textbf{Model} & \textbf{\dsOne} & \textbf{\dsSix} & \textbf{\stThree} & \textbf{\qwCodeOne} & \textbf{\qwCodeSeven} & \textbf{\qwThreeZ} & \textbf{\qwThreeOne} & \textbf{\qwThreeFour} & \textbf{\qwThreeEight} \\
\midrule
$\bm{\rho}$ & 0.4728 & 0.4595 & 0.4620 & 0.4906 & 0.4755 & 0.5249 & 0.5028 & 0.4936 & 0.4877 \\
\bottomrule
\end{tabular}
}
\end{table}

%% file: sections/approach.tex
\section{Approach}\label{sec:approach}
We now introduce \textbf{\app}, an uncertainty-guided adaptive decoding framework for LLM-based code generation. \app integrates a token-level \textit{\textbf{pause-then-rerank}} mechanism into the decoding process of the target LLM, drawing on insights from our empirical analysis.

\subsection{Overview}

\input{figures/workflow}

As illustrated in Figure~\ref{fig:workflow}, \app is an adaptive (dynamic) decoding framework: at inference time it continuously monitors per-step uncertainty and dynamically decides whether to intervene, rather than applying a uniform policy across all steps or only at line-start positions. Given a target model \textbf{LM}, at each decoding step \(t\) the model produces a next-token distribution \(\bm{p}_t\). \app computes the Shannon entropy \(H(\bm{p}_t)\) and compares it against a model-specific threshold \(\tau^{\bm{LM}}\). If \(H(\bm{p}_t) \le \tau^{\bm{LM}}\), the framework proceeds with the base decoding policy (e.g., greedy, top-\(k\), or top-\(p\) sampling). If \(H(\bm{p}_t) > \tau^{\bm{LM}}\), \app pauses normal decoding and triggers a targeted reranking routine over the top-\(K\) candidates. During reranking, each candidate undergoes a short lookahead of up to \(N\) steps to estimate the quality of its future trajectory, and the candidate with the highest score is selected as the next token. This per-step decision loop repeats until an end-of-sequence (EOS) token is generated.

The adaptive nature of \app has three practical consequences. First, the intervention is selective and data-driven: only high-uncertainty steps incur additional computation, which limits overhead compared to always-on strategies. Second, the threshold \(\tau^{\bm{LM}}\) is learned in a preparation stage (we describe the learning procedure in Sec.~\ref{sec:pause}) so that the pause decision is model-aware rather than hand-tuned. Third, the framework supports flexible reranking configurations (e.g., adjustable lookahead length \(N\) and lookahead width \(K\)), allowing trade-offs between accuracy and latency to be tuned for different deployment constraints.

\app is designed to be broadly applicable. Since the pause-then-rerank mechanism is triggered purely by uncertainty signals (entropy), it can be seamlessly integrated with various decoding strategies, including sampling. This flexibility underscores the generality of \app: rather than being tied to a specific decoding style, it acts as a plug-in framework that can enhance both deterministic (e.g., greedy) and stochastic (e.g., sampling) generation. Unless otherwise specified, we use \textbf{greedy decoding as the base policy} throughout the remainder of this paper.

The details of the pause checking and lookahead reranking are described in the following sections.

\subsection{Entropy‐based Pause Checking}\label{sec:pause}

We use the Shannon entropy $H(\bm{p}_t)$ over the probability distribution $\bm{p}_t$ predicted by the model \textbf{LM} to measure the token uncertainty. When $H$ exceeds a threshold, \app pauses the decoding process. As revealed in the empirical study, a fixed entropy threshold may not generalize well across models with different output distributions. Therefore, we propose a data-driven method to learn a model-specific threshold $\tau^{\bm{LM}}$ using logistic regression.

\subsubsection{Data Collection}\label{sec:training-data}
We cast threshold learning as a binary classification task. Each decoding step is a sample, where the input feature is the entropy $H(\bm{p}_t)$ and the label is 1 if the ground-truth token is the top-1 prediction, and 0 otherwise. Training data is generated via ground-truth–guided next-token prediction on BigCodeBench~\cite{zhuo2024bigcodebench}. To strictly mitigate data contamination, we verified semantic similarity between BigCodeBench and our evaluation sets using SBERT~\cite{sbert}. Manual review of all problem pairs exceeding a 0.7 cosine similarity threshold (maximum observed was 0.75) confirmed zero overlapping duplicates. Through this process, we collect approximately 100K--–160K positive samples and 16K--23K negative samples per model (e.g., 161K positive and 19K negative for \dsSix).

We then balance the dataset by randomly downsampling the positive samples and selecting an equal number of negative samples, which is a common technique for handling imbalanced classification tasks~\cite{lemaitre2016imbalancedlearnpythontoolboxtackle}. This choice is driven by the nature of logic drift in code generation: once a drift point occurs, it typically leads to irreversible semantic errors. Therefore, it is essential to maximize the classifier’s ability to detect negative cases, even if they are relatively scarce. By constructing a balanced dataset, we prevent the model from being biased toward the majority class and enable it to more effectively learn the decision boundary for identifying uncertain decoding steps. This strategy enhances the classifier’s sensitivity to potential drift points, thereby improving the overall reliability of the entropy-based triggering mechanism.

\subsubsection{Regression Model Training}
We train a logistic regression model to predict whether the ground truth token is the top-1 prediction, using only the entropy value as input. The model estimates the probability of a positive label using the sigmoid function:
$$
P(y=1 \mid H^{(i)}) = \frac{1}{1 + \exp(-(\beta_0 + \beta_1 H^{(i)}))},
$$
where $H^{(i)}$ denotes the entropy of the $i$-th sample, and $\beta_0$, $\beta_1$ are the model parameters to be learned. These parameters are learned by minimizing the binary cross-entropy loss through maximum likelihood estimation. Note that, since the training data is collected using the target \textbf{LM}, the trained logistic regression model is specific to that \textbf{LM}.

\subsubsection{Threshold Estimation}
To estimate a model-specific entropy threshold $\tau^{\bm{LM}}$, we use the trained logistic regression model to perform binary classification on the validation set, where the task is to predict whether the top-1 token matches the ground-truth token. As the logistic regression model outputs probabilities, we first identify the optimal probability threshold that maximizes classification accuracy, and then convert this probability threshold into an equivalent entropy threshold.

Specifically, we conduct a grid search over probability thresholds in the range [0.01, 0.99], using a step size of 0.01. For each candidate threshold, a sample is classified as positive if its predicted probability exceeds the threshold, and negative otherwise. The threshold that achieves the highest classification accuracy on the validation set is selected as the optimal probability threshold, denoted by $p^*$. We then convert this optimal probability threshold into the corresponding entropy threshold using the parameters of the trained logistic regression model. Given the intercept $\beta_0$ and the coefficient $\beta_1$, the entropy threshold $\tau^{\bm{LM}}$ is computed using the log-odds transformation:
\[
\tau^{\bm{LM}} = \frac{\log\left(\frac{p^*}{1 - p^*}\right) - \beta_0}{\beta_1}.
\]
Entropy values below $\tau^{\bm{LM}}$ are classified as positive (i.e., the model is likely to correctly predict the top-1 token), while those above are classified as negative. This entropy threshold is then used as the final decision boundary for the target language model during decoding.

\subsubsection{Pause Triggering}
During decoding, \app uses the learned threshold $\tau^{\bm{LM}}$ to decide when to trigger candidate reranking. If the entropy at a decoding step exceeds $\tau^{\bm{LM}}$, it indicates high uncertainty, suggesting that the top-1 prediction may be incorrect. In such cases, the decoding process is paused, and the top candidates are reranked to improve accuracy.

\subsection{Lookahead-based Token Reranking}\label{sec:lookahead}
\app employs a lookahead re-scoring strategy for token reranking, enhancing selection by assessing the future impact of each candidate token at every decoding step. Inspired by A* search heuristics and prior work on constrained generation~\cite{lu2021neurologic}, this method extends token probability evaluation beyond immediate next-token predictions to consider longer-term sequence quality. The process is governed by two hyperparameters: the \textit{lookahead beam size} ($B$), which determines the number of candidate tokens considered, and the \textit{lookahead length} ($L$), which sets the length of the future token sequence evaluated.

\subsubsection{Lookahead Process}
At decoding step $t$, the model outputs a probability distribution $\bm{p}_t$ over the vocabulary for the next token. From this distribution, we select the top-$B$ tokens to form the candidate set $\mathcal{C}_t = \{c_t^{(1)}, c_t^{(2)}, \dots, c_t^{(B)}\}$.

For each candidate token $c_t^{(k)}$, we construct a lookahead trajectory of length $L$ by appending $L$ tokens generated via greedy decoding, conditioned on the current input and previously generated tokens. This results in the trajectory $\mathbf{T}_t^{(k)}$:
$$
\mathbf{T}_t^{(k)} = (c_t^{(k)}, \hat{c}_{t+1}^{(k)}, \hat{c}_{t+2}^{(k)}, \dots, \hat{c}_{t+L}^{(k)}),
$$
where each $\hat{c}_{t+i}^{(k)}$ is the most probable token predicted at step $t+i$, given the prefix ending with $c_t^{(k)}$.

\textbf{Trajectory Scoring.}
We evaluate each trajectory by computing the average log-probability of its tokens:
$$
\text{score}(c_t^{(k)}) = \frac{1}{L+1} \left[ \log P(c_t^{(k)}) + \sum_{i=1}^{L} \log P(\hat{c}_{t+i}^{(k)}) \right],
$$
where $P(\cdot)$ denotes the model's predicted probability for a token. Because we apply an early stopping mechanism, different candidates may result in varying lookahead lengths. To ensure a fair comparison across such candidates, we normalize the total log-probability by computing the geometric mean of token probabilities in the above equation, rather than relying solely on log-likelihood. This normalization mitigates the bias against longer sequences, which tend to accumulate lower joint probabilities. The resulting score captures the model's overall confidence in the full trajectory, balancing immediate token likelihood with the coherence of future predictions.

\textbf{Token Selection.}
The candidate with the highest trajectory score is selected as the output token at step $t$:
$$
c^*_t = \arg\max_{c_t^{(k)} \in \mathcal{C}_t} \text{score}(c_t^{(k)}).
$$
This process is repeated at each decoding step, ensuring that token selection is guided not just by immediate probabilities, but by the expected quality of the downstream sequence.

This lookahead mechanism improves generation quality by favoring tokens that lead to fluent and coherent continuations, which is particularly important in structured domains like code generation. The hyperparameters $B$ and $L$ control the trade-off between computational cost and generation quality: larger values offer better foresight at the expense of increased inference time.

\subsubsection{Lookahead Beam Size Selection}
To determine an appropriate value for $B$, we analyze the training data collected in Section~\ref{sec:training-data}. Specifically, we examine the average rank of the ground truth token at decoding steps where the Shannon entropy is either above or below the threshold $\tau^{\bm{LM}}$, which is learned via logistic regression. After filtering out samples where the ground truth token's rank exceeds 20, we observe a clear distinction: when the entropy is below or equal to the threshold, the average rank typically falls between 1.1 and 1.3, indicating high model confidence. In contrast, when the entropy exceeds the threshold—thus triggering a pause—the average rank increases to between 2.6 and 3.4. 

These findings support the use of a small beam size (e.g., $B = 3$) to achieve a practical trade-off between candidate coverage and computational efficiency.

\subsubsection{Lookahead Length Selection}
We adopt a lookahead length parameter $L$ when performing trajectory scoring. At each reranking step we simulate up to $L$ future tokens for each candidate to estimate the quality of its continuation. The parameter $L$ controls the trade-off between accuracy and efficiency: a larger $L$ provides a more reliable estimate of long-term trajectory quality but incurs higher computation cost, whereas a smaller $L$ reduces overhead but may fail to capture meaningful differences among candidates.

In practice, we find that moderate values of $L$ (e.g., $L=5$) are sufficient to balance these concerns, since code generation typically exhibits strong local dependencies and short-horizon lookahead already provides useful signals. We analyze the effect of varying $L$ through experiments in Section~\ref{sec:evaluation/RQ4}.

%% file: figures/workflow.tex
\begin{figure}[t]
    \centering
    \includegraphics[width=1\columnwidth]{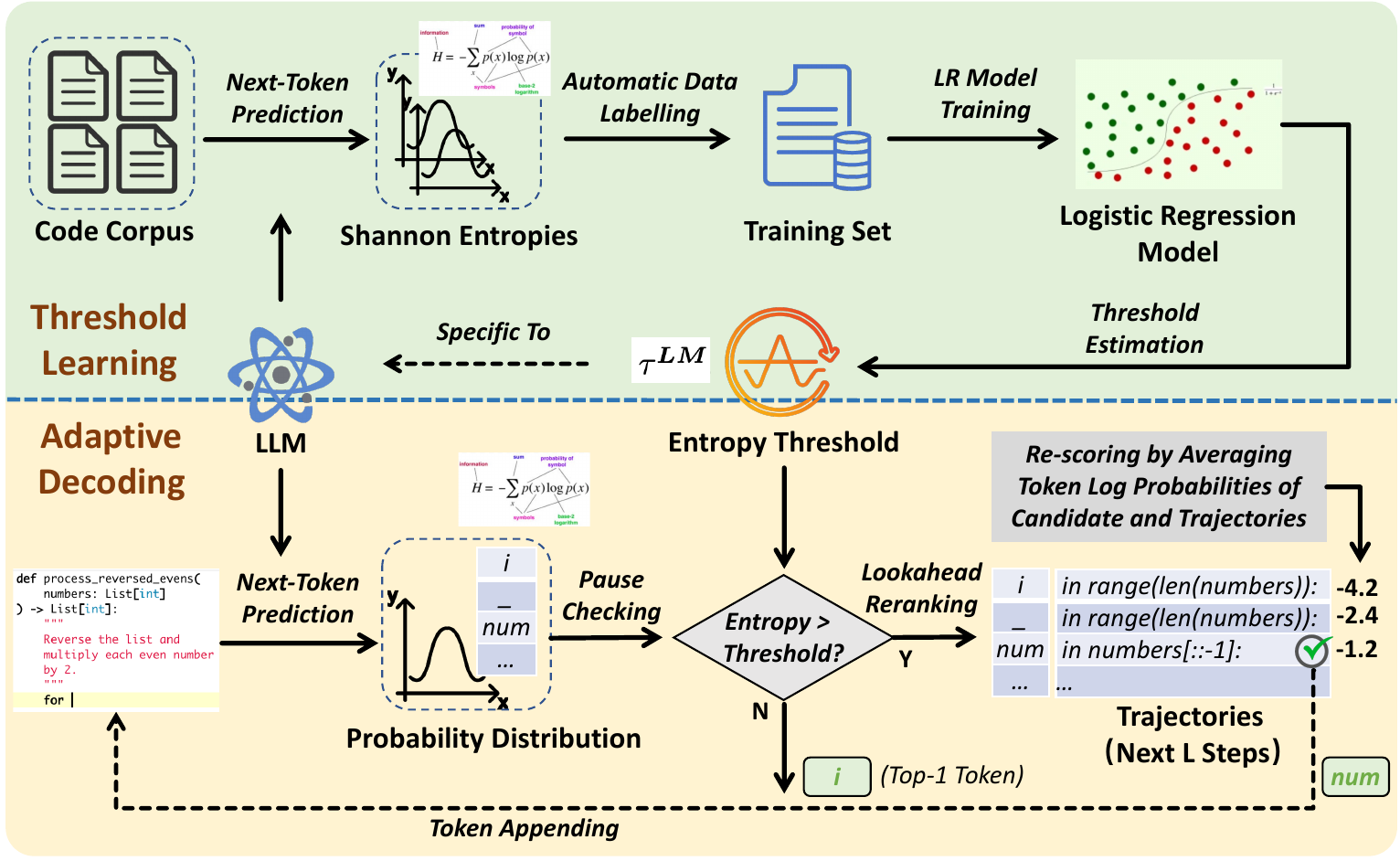}
    \Description{}
    \caption{Approach Overview of \app}
    \label{fig:workflow}
\end{figure}

%% file: sections/evaluation.tex
\section{Evaluation}\label{sec:evaluation}

We conduct extensive experiments to evaluate the effectiveness of \app. Specifically, we aim to answer the following research questions:

\begin{itemize}
  \item \textbf{RQ1 (Effectiveness):} How effective is \app in improving the functional correctness of generated code?
  \item \textbf{RQ2 (Efficiency):} How efficient is \app in terms of computational cost?
  \item \textbf{RQ3 (Ablation Study):} How effective is the logistic regression–based threshold learning component, and can the learned thresholds better guide practical decoding?
  \item \textbf{RQ4 (Parameter Sensitivity):} How does \app perform under different values of the lookahead length $L$?
\end{itemize}

\subsection{Experimental Setup}
Our experimental setup is as follows.

\textbf{Datasets.} We evaluate our method on three widely used code generation benchmarks:

\begin{itemize}
    \item \textbf{HumanEval+}~\cite{evalplus}: An extended version of HumanEval\cite{Humaneval} with test cases increased by a factor of 81 for 164 Python problems, enabling rigorous functional correctness evaluation.
    \item \textbf{MBPP+}~\cite{evalplus}: A refined subset of 378 MBPP\cite{MBPP} problems with 35 times larger unit tests for stricter evaluation.
    \item \textbf{DevEval}~\cite{Deveval}: A repository-level benchmark aligned with real-world scenarios. We randomly sample 238 problems from its 1,874 samples, ensuring 90\% confidence within a ±5\% margin.
\end{itemize}

\textbf{Baselines.} We compare \app against three decoding strategies:

\begin{itemize}
    \item \textbf{Greedy Decoding}~\cite{Radford2019LanguageMA}: Selects the highest-probability token at each step; computationally efficient but myopic.
    \item \textbf{Beam Search}~\cite{beamsearch}: Maintains the top-$k$ sequences. We set beam size $B=3$ to match \app's candidate pool, isolating the benefits of uncertainty-guided reranking.
    \item \textbf{AdapT}~\cite{AdapT}: Adjusts sampling temperature for line-start (high-uncertainty) tokens. We empirically set $a=0.5$ and $b=0.05$ (for other tokens) for optimal benchmark performance.
\end{itemize}

We also considered \textbf{Uncert-CoT}~\cite{UnCert-CoT}, another uncertainty-guided decoding method originally proposed for chain-of-thought reasoning tasks. However, we do not include it in our experimental comparison since no official implementation is available and the original paper omits several crucial details necessary for faithful reproduction.

\subsection{RQ1: Effectiveness in Improving Functional Correctness}
Table~\ref{tab:pass1_comp} presents the Pass@1 performance of four decoding strategies—Greedy, Beam Search ($B=3$), AdapT, and \app—across nine LLMs evaluated on the HumanEval+, MBPP+, and DevEval datasets.

\input{tables/pass1_comp}

\textbf{Comparison with Greedy Search.}
Across all three benchmarks, \app consistently outperforms Greedy decoding. 
On HumanEval+, \app improves Pass@1 by an average of \textbf{+3.59\%}, with gains as high as +6.09\% for \qwThreeEight and +5.49\% for \qwThreeZ. 
On MBPP+, the improvement is even more pronounced: \app yields an average gain of \textbf{+7.61\%}, with particularly significant performance boosts observed for \qwThreeZ (+20.90\%) and \qwThreeEight (+17.72\%).
On DevEval, \app demonstrates robust effectiveness, yielding meaningful improvements for all evaluated models (\textbf{+4.53\%} on average). Particularly, \qwThreeFour and \stThree show strong gains of +7.57\% and +7.14\%, respectively. 
Unlike the mixed results sometimes observed with other strategies, \app achieves positive improvements across all models on DevEval, highlighting its ability to handle repository-level contexts where long-range logic consistency is crucial.

\textbf{Comparison with Beam Search.}
A critical question is whether the gains of \app stem merely from searching multiple candidate tokens or from our uncertainty-guided strategy. 
Comparing \app with standard Beam Search (using the same beam width $B=3$), we observe that \app consistently achieves superior performance. 
While Beam Search provides a baseline improvement over Greedy (averaging +3.05\% on HumanEval+ and +4.00\% on MBPP+), \app surpasses it by further margins. 
The advantage of \app is particularly evident on the more challenging MBPP+ and DevEval benchmarks. For instance, on DevEval, Beam Search yields a modest average gain of +1.45\%, whereas \app achieves \textbf{+4.53\%}, nearly tripling the improvement. 
This indicates that blindly searching for top-$k$ continuations (as in Beam Search) is less effective than \app's approach of triggering lookahead only at high-uncertainty points, which allows for more targeted and semantically meaningful corrections.

\textbf{Comparison with AdapT.}
When compared to AdapT, \app shows clear advantages in both stability and magnitude of improvement. 
While AdapT offers marginal improvements over Greedy decoding (e.g., +1.42\% on HumanEval+ and +0.59\% on MBPP+), it exhibits instability, leading to performance drops on DevEval for several models (average -0.47\%). 
In contrast, \app consistently achieves larger gains, surpassing AdapT by wide margins across nearly all models and datasets. 
For example, on MBPP+, \app exceeds AdapT by approximately \textbf{+7\%} on average. 
These results confirm that utilizing Shannon entropy to guide lookahead is a more reliable strategy than the heuristic-based temperature adjustment used in AdapT.

\summary{
Overall, the results indicate that \app is highly effective in improving the functional correctness of generated code. It consistently outperforms Greedy decoding across diverse models and benchmarks. More importantly, it surpasses Beam Search, particularly on complex tasks (MBPP+ and DevEval), proving that uncertainty-guided intervention is superior to blind search. Compared with AdapT, \app demonstrates significantly more robust and scalable gains.
}

\subsection{RQ2: Efficiency in Computational Cost}
We evaluate the efficiency of \app in terms of computational cost on the HumanEval+, MBPP+, and DevEval datasets. We analyze two key metrics: the \textit{pause rate} and \textit{prediction latency}.

\textbf{Pause Rate.}  
Table~\ref{tab:lookahead_ratio} presents the pause rates across different models and datasets. The results show that \app triggers the reranking mechanism sparingly, with pause rates ranging from 0.85\% to 12.17\%, averaging approximately \textbf{6.75\%} across all configurations.
This low activation frequency indicates that the entropy-based threshold successfully identifies a small subset of critical, high-uncertainty steps. Unlike global search strategies, \app concentrates computational effort only where necessary. Notably, even for the largest models, the pause rate generally remains below 10\%, demonstrating that our selective triggering mechanism scales efficiently without incurring the cost of constant lookahead.

\input{tables/lookahead_ratio}

\textbf{Latency.}  
Table~\ref{tab:runtime_comp} reports the average generation time per problem for Greedy, Beam Search ($B=3$), AdapT, and \app.

\textit{Comparison with Greedy.} 
Overall, \app introduces moderate additional latency compared to Greedy decoding (e.g., an average increase of +0.75s on HumanEval+ and +0.32s on MBPP+). This overhead is expected due to the lookahead computations. However, considering the substantial accuracy gains reported in RQ1, this trade-off is highly favorable.

Notably, for certain models (e.g., \qwThreeZ and \qwThreeOne on HumanEval+), \app actually exhibits lower latency than Greedy decoding. This counter-intuitive result occurs because \app prevents the model from drifting into erroneous, low-quality trajectories—such as repetitive loops or verbose, nonsensical outputs—that often plague Greedy decoding at high-uncertainty points. By correcting the generation path early via lookahead, \app produces more concise and correct solutions, thereby reducing the total number of generated tokens and effectively offsetting the computational cost of reranking.

\textit{Comparison with Beam Search.} 
A critical finding is that \app is significantly more efficient than Beam Search, despite outperforming it in accuracy. 
As shown in Table~\ref{tab:runtime_comp}, Beam Search introduces substantial latency overhead due to maintaining multiple active beams at every step. For instance, on the repository-level DevEval benchmark, Beam Search increases the average generation time by \textbf{+24.62s} compared to Greedy. In contrast, \app increases it by only \textbf{+7.47s}. 
This efficiency stems from our \textbf{selective pausing strategy}: by triggering the lookahead mechanism only at high-uncertainty steps rather than globally, \app avoids wasted computation on trivial tokens.

\textit{Comparison with AdapT.} 
AdapT exhibits mixed runtime behavior. In simpler scenarios (e.g., MBPP+), it adds negligible overhead (+0.10s) or even reduces time slightly for specific models by promoting shorter sequences via sampling. However, \app provides a more consistent balance: while slightly slower than AdapT, it delivers far more robust functional correctness (as shown in RQ1), justifying the marginal increase in runtime.

\input{tables/runtime_comp}

\summary{
\app maintains high efficiency by selectively triggering lookahead at a low rate (average $\approx$ 6.75\%), effectively filtering out low-utility computations. Crucially, compared to Beam Search, \app incurs significantly lower additional overhead while achieving superior output quality across all benchmarks. This confirms that uncertainty-guided lookahead represents a far more efficient allocation of inference resources than blind search.
}

\subsection{RQ3: Ablation Study}  
We perform an ablation study to evaluate the key components of \app. In particular, we investigate whether the learned entropy threshold $\tau^{\bm{LM}}$ serves as an effective signal, both in terms of predictive ability (as a classifier) and in practical decoding performance.

\textbf{Predictive Effectiveness of Learned Entropy Thresholds.}
Table~\ref{tab:LR_evaluation} summarizes the performance of the logistic regression classifier trained with learned entropy thresholds ($\tau^{\bm{LM}}$). 
Overall, the classifier demonstrates robust predictive capability across all models, achieving an average F1 score of 0.9351 and an AUC of 0.9116.
The highest performance is observed on \dsSix, with an accuracy of 91.05\% and an F1 score of 0.9512. Even for smaller models, the classifier remains reliable; for instance, \stThree achieves an F1 score of 0.9475.
These high metrics confirm that Shannon entropy, when calibrated via our learning process, serves as a high-fidelity signal for distinguishing between correct and incorrect token predictions.

\input{tables/LR_evaluation}

\textbf{Practical Effectiveness of Learned Entropy Thresholds.}
To validate the necessity of learning the threshold rather than using a heuristic constant, we compare the decoding performance of models using the learned threshold versus a fixed threshold. The fixed threshold is set to 1.2. 
Table~\ref{tab:threshold_comparison} reports the Pass@1 results on HumanEval+.

The experiments demonstrate that using the learned thresholds yields consistent improvements over the fixed threshold baseline across all evaluated models. 
The average improvement is \textbf{+1.76\%}, with gains ranging from +0.61\% to +4.88\%.
Specifically, while most models (e.g., \dsOne, \qwCodeSeven, \qwThreeEight) show stable moderate gains between 1.2\% and 1.8\%, distinctively larger boosts are observed for \qwThreeOne (+3.05\%) and \qwThreeFour (+4.88\%).
This indicates that a fixed threshold of 1.2 is often suboptimal. By calibrating the threshold to each model's specific uncertainty distribution, \app ensures robust performance gains, effectively unlocking the model's potential regardless of its specific architecture or scale.

\input{tables/threshold_comparison}

\summary{
The learned entropy threshold $\tau^{\bm{LM}}$ is both theoretically sound and practically effective. It reliably predicts token risks and consistently outperforms a fixed-threshold baseline across all models, delivering an average Pass@1 improvement of +1.76\% and maximizing the effectiveness of the uncertainty-guided intervention.
}

\subsection{RQ4: Parameter Sensitivity}\label{sec:evaluation/RQ4}
To investigate the impact of key hyperparameters on decoding performance, we conduct sensitivity analyses on the lookahead length $L$ and the learned entropy threshold $\tau^{LM}$. All experiments in this section use \dsOne as the backbone model evaluated on HumanEval+.

\textbf{Sensitivity to Lookahead Length ($L$).} 
We vary $L$ from 2 to 9 and report Pass@1 results. Figure~\ref{fig:L_analysis} illustrates the trend.

\input{figures/L_analysis}

The results show that Pass@1 increases steadily with larger $L$ up to $L=5$, at which point it reaches the peak performance of \textbf{65.24\%}. 
Beyond $L=5$, the performance does not continue to rise; instead, it exhibits slight fluctuations, hovering between 63.41\% and 64.63\%. 
This behavior reflects the trade-off between local guidance and global noise: while shorter lookaheads may fail to capture sufficient future context, excessively long lookaheads ($L > 5$) may introduce noise from uncertainty accumulation, diluting the relevance of immediate local signals. 

Crucially, across all values of $L$, \app consistently outperforms the Greedy baseline (62.20\%). Even at the minimal setting of $L=2$, \app achieves 62.80\%, reducing to a safe margin of +0.60\%. This demonstrates that our approach is robust to hyperparameter choices and provides stable improvements.

\textbf{Sensitivity to Threshold Variations ($\tau^{LM}$).} 
As illustrated in Table~\ref{tab:sensitivity}, we evaluate the Pass@1 accuracy by shifting the optimal learned threshold ($\tau^{LM} \approx 0.9850$) across various intervals. \app demonstrates high robustness to minor fluctuations. Within the $[\tau^{LM}-0.1, \tau^{LM}+0.1]$ range, Pass@1 remains highly stable and consistently outperforms the greedy baseline. 

\input{tables/sensitivity}

Visible performance degradation occurs only under drastic shifts ($\pm 0.5$): excessively low thresholds introduce noise via over-intervention, while excessively high thresholds become too conservative. This confirms that $\tau^{LM}$ is not a fragile hyperparameter requiring extreme precision.

\summary{
Both key hyperparameters exhibit high robustness. The lookahead length $L$ shows a non-monotonic effect, with moderate values (e.g., $L=5$) maximizing the benefits of lookahead guidance. The learned threshold $\tau^{LM}$ is highly stable against minor fluctuations within $\pm 0.1$. Notably, \app consistently surpasses baseline performance across varying hyperparameter settings.
}

%% file: tables/pass1_comp.tex
\newcommand{\resBase}[1]{
  \begin{tabular}{@{}c@{}}#1 \\[-5pt] \phantom{\scriptsize{(+0.00)}}\end{tabular}%
}

\newcommand{\resInc}[2]{
  \begin{tabular}{@{}c@{}}#1 \\[-5pt] \scriptsize{\color{black!65}(#2)}\end{tabular}%
}

\begin{table}[htbp]
  \centering
  \caption{Pass@1 (\%) comparison of Greedy, Beam Search, AdapT, and \app on HumanEval+, MBPP+, and DevEval. Values in parentheses indicate the change relative to Greedy.}
  \label{tab:pass1_comp}
  
  \renewcommand{\arraystretch}{1.25} 
  \setlength{\tabcolsep}{3pt}
  
  \resizebox{\linewidth}{!}{
    \begin{tabular}{l|cccc|cccc|cccc}
      \toprule
      \multirow{2}{*}{\textbf{Model}} & \multicolumn{4}{c|}{\textbf{HumanEval+}} & \multicolumn{4}{c|}{\textbf{MBPP+}} & \multicolumn{4}{c}{\textbf{DevEval}} \\
      \cmidrule(lr){2-5} \cmidrule(lr){6-9} \cmidrule(lr){10-13}
      & Greedy & Beam & AdapT & \textbf{\app} & Greedy & Beam & AdapT & \textbf{\app} & Greedy & Beam & AdapT & \textbf{\app} \\
      \midrule
      
      \resBase{\dsOne} & 
      \resBase{62.20} & \resInc{62.20}{+0.00} & \resInc{63.41}{+1.21} & \resInc{65.24}{+3.04} & 
      \resBase{52.12} & \resInc{51.59}{-0.53} & \resInc{52.65}{+0.53} & \resInc{53.17}{+1.05} & 
      \resBase{14.29} & \resInc{18.91}{+4.62} & \resInc{12.61}{-1.68} & \resInc{18.49}{+4.20} \\

      \resBase{\dsSix} & 
      \resBase{71.34} & \resInc{76.83}{+5.49} & \resInc{72.56}{+1.22} & \resInc{73.78}{+2.44} & 
      \resBase{65.61} & \resInc{65.87}{+0.26} & \resInc{66.93}{+1.32} & \resInc{67.72}{+2.11} & 
      \resBase{25.21} & \resInc{28.57}{+3.36} & \resInc{26.47}{+1.26} & \resInc{26.47}{+1.26} \\

      \resBase{\stThree} & 
      \resBase{48.78} & \resInc{50.61}{+1.83}& \resInc{51.83}{+3.05} & \resInc{50.61}{+1.83} & 
      \resBase{38.89} & \resInc{45.77}{+6.88} & \resInc{39.42}{+0.53} & \resInc{46.03}{+7.14} & 
      \resBase{15.55} & \resInc{13.03}{-2.52} & \resInc{13.03}{-2.52} & \resInc{22.69}{+7.14} \\

      \resBase{\qwCodeOne} & 
      \resBase{64.02} & \resInc{68.90}{+4.88} & \resInc{64.02}{+0.00} & \resInc{67.07}{+3.05} & 
      \resBase{59.79} & \resInc{61.64}{+1.85} & \resInc{60.32}{+0.53} & \resInc{62.17}{+2.38} & 
      \resBase{10.08} & \resInc{11.34}{+1.26} & \resInc{10.50}{+0.42} & \resInc{14.71}{+4.63} \\

      \resBase{\qwCodeSeven} & 
      \resBase{84.15} & \resInc{83.54}{-0.61} & \resInc{87.20}{+3.05} & \resInc{87.20}{+3.05} & 
      \resBase{71.16} & \resInc{71.96}{+0.80} & \resInc{72.22}{+1.06} & \resInc{72.22}{+1.06} & 
      \resBase{18.49} & \resInc{14.71}{-3.78} & \resInc{18.07}{-0.42} & \resInc{23.11}{+4.62} \\

      \resBase{\qwThreeZ} & 
      \resBase{17.07} & \resInc{24.39}{+7.32}& \resInc{18.29}{+1.22} & \resInc{22.56}{+5.49} & 
      \resBase{9.52}  & \resInc{21.16}{+11.64}& \resInc{11.11}{+1.59} & \resInc{30.42}{+20.90}& 
      \resBase{2.94}  & \resInc{5.88}{+2.94} & \resInc{2.52}{-0.42}  & \resInc{7.56}{+4.62} \\

      \resBase{\qwThreeOne} & 
      \resBase{40.24} & \resInc{41.46}{+1.22} & \resInc{42.07}{+1.83} & \resInc{44.51}{+4.27} & 
      \resBase{37.57} & \resInc{43.39}{+5.82} & \resInc{36.51}{-1.06} & \resInc{42.86}{+5.29} & 
      \resBase{7.98}  & \resInc{9.24}{+1.26} & \resInc{7.56}{-0.42}  & \resInc{11.76}{+3.78} \\

      \resBase{\qwThreeFour} & 
      \resBase{53.05} & \resInc{57.32}{+4.27} & \resInc{53.66}{+0.61} & \resInc{56.10}{+3.05} & 
      \resBase{40.48} & \resInc{42.86}{+2.38}& \resInc{41.01}{+0.53} & \resInc{51.32}{+10.84}& 
      \resBase{9.24}  & \resInc{13.45}{+4.21} & \resInc{9.24}{+0.00}  & \resInc{16.81}{+7.57} \\

      \resBase{\qwThreeEight} & 
      \resBase{56.71} & \resInc{59.76}{+3.05}& \resInc{57.32}{+0.61} & \resInc{62.80}{+6.09} & 
      \resBase{29.37} & \resInc{36.24}{+6.87}& \resInc{29.63}{+0.26} & \resInc{47.09}{+17.72}& 
      \resBase{4.62}  & \resInc{6.30}{+1.68} & \resInc{4.20}{-0.42}  & \resInc{7.56}{+2.94} \\
      
      \midrule
      \textbf{AVG.} $\Delta$ & -- & \textbf{+3.05}& \textbf{+1.42} & \textbf{+3.59} & -- & \textbf{+4.00}& \textbf{+0.59} & \textbf{+7.61} & -- & \textbf{+1.45} & \textbf{-0.47} & \textbf{+4.53} \\
      \bottomrule
    \end{tabular}
  }
\end{table}

%% file: tables/lookahead_ratio.tex
\begin{table}[htbp]
    \centering
    \small
    \caption{Pause ratio of the models on HumanEval+, MBPP+, and DevEval.}
    \label{tab:lookahead_ratio}
    \renewcommand{\arraystretch}{1.25} 
    \setlength{\tabcolsep}{2pt}
  
    \resizebox{\linewidth}{!}{
    \begin{tabular}{lccccccccc}
        \toprule
                          & \textbf{\dsOne} & \textbf{\dsSix} & \textbf{\stThree}  & \textbf{\qwCodeOne} & \textbf{\qwCodeSeven} & \textbf{\qwThreeZ} & \textbf{\qwThreeOne} & \textbf{\qwThreeFour} & \textbf{\qwThreeEight} \\
        \midrule
        HumanEval+        & 2.81\%           & 2.33\%& 4.80\%          & 1.48\%&0.85\%& 6.11\%           & 11.17\%          & 7.82\%         & 5.98\%         \\
        MBPP+             & 8.49\%& 7.46\%           & 6.52\%          & 6.24\%&4.43\%& 8.50\%           & 12.17\%          & 7.11\%         & 8.33\%         \\
        DevEval           & 7.75\%& 6.29\%& 6.59\%          & 6.95\%&9.70\%& 6.16\%           & 11.47\%           & 9.69\%         & 5.04\%         \\
        \midrule
        \textbf{AVG.} $\Delta$ & 6.35\%& 5.36\%& 5.97\%  & 4.89\%&4.99\%& 6.92\% & 11.60\% & 8.21\% & 6.45\% \\
        \bottomrule
    \end{tabular}
    }
\end{table}

%% file: tables/runtime_comp.tex
\begin{table}[ht]
\centering
\renewcommand{\arraystretch}{1.1}
\caption{Comparison of the average generation time per problem (in seconds) by Greedy, Beam Search, AdapT, \app on HumanEval+, MBPP+, DevEval.}
\label{tab:runtime_comp}

\renewcommand{\arraystretch}{1.25}
\setlength{\tabcolsep}{3pt}

\resizebox{\textwidth}{!}{
\begin{tabular}{l|cccc|cccc|cccc}
\toprule
\multirow{2}{*}{\textbf{Model}} & \multicolumn{4}{c|}{\textbf{HumanEval+}} & \multicolumn{4}{c|}{\textbf{MBPP+}} & \multicolumn{4}{c}{\textbf{DevEval}} \\
\cmidrule(lr){2-5} \cmidrule(lr){6-9} \cmidrule(lr){10-13}
& Greedy & Beam & AdapT & \app & Greedy & Beam & AdapT & \app & Greedy & Beam & AdapT & \app \\
\midrule
\dsOne         & 2.94 & 7.01& 2.72& 4.56 & 1.46& 1.91& 1.47& 1.68& 4.03& 14.59& 4.24& 7.66\\
\dsSix         & 9.55& 25.60& 13.59& 11.87& 3.21& 7.51& 3.13& 4.81& 20.56& 53.52& 23.72& 32.91\\
\stThree       & 1.71& 3.45& 2.72& 2.06 & 0.60& 1.01& 0.84& 0.77& 6.92& 34.40& 8.33& 14.90\\
 \qwCodeOne& 5.29& 9.39& 4.18& 5.82& 1.78& 2.24& 1.55& 2.24& 7.91& 19.02& 7.88&10.12\\
 \qwCodeSeven& 12.24& 38.22& 13.40& 14.38& 3.94& 10.12& 3.90& 6.02& 12.12& 26.51& 13.71&22.19\\
\qwThreeZ      & 6.34& 8.94& 5.18& 5.03& 5.94& 9.44& 6.14 & 3.89& 19.49& 23.79& 17.86& 21.84\\
\qwThreeOne    & 8.23& 9.98& 6.03& 5.29& 2.13& 3.17& 2.54 & 2.28 & 22.21& 46.98& 20.56& 29.53\\
\qwThreeFour   & 5.41& 9.13& 7.29& 8.10& 3.49& 5.02& 3.76 & 3.37 & 29.57& 80.24& 31.94& 42.78\\
\qwThreeEight  & 5.62& 11.76& 5.63& 6.99& 3.88& 5.19& 4.03& 4.21& 61.14& 106.48& 63.53& 69.23\\
\midrule
\textbf{AVG.} $\Delta$ & -- & \textbf{+7.35}& \textbf{+0.38}& \textbf{+0.75}& -- & \textbf{+2.13}& \textbf{+0.10}& \textbf{+0.32}& -- & \textbf{+24.62}& \textbf{+0.87}& \textbf{+7.47}\\
\bottomrule
\end{tabular}
}
\end{table}

%% file: tables/LR_evaluation.tex
\begin{table}[ht]
\centering
\renewcommand{\arraystretch}{1.2}
\caption{Evaluation of logistic regression models trained to predict whether the ground truth token is the top-1 prediction. The learned $\bm{\tau}^{\bm{LM}}$ denotes the entropy threshold selected via logistic regression.}
\label{tab:LR_evaluation}
\begin{tabular}{lcccccc}
\toprule
\textbf{Model} & \textbf{Accuracy} & \textbf{Precision} & \textbf{Recall} & \textbf{F1 Score} & \textbf{AUC} & \textbf{Learned $\bm{\tau}^{\bm{LM}}$} \\
\midrule
\dsOne   & 0.8994 & 0.9251 & 0.9640 & 0.9442 & 0.9268 & 0.9850 \\
\dsSix   & 0.9105 & 0.9276 & 0.9761 & 0.9512 & 0.9305 & 1.1338 \\
\stThree     & 0.9038 & 0.9188 & 0.9780 & 0.9475 & 0.9222 & 1.1005 \\
 \qwCodeOne& 0.8824& 0.9055& 0.9638& 0.9337& 0.9085&1.2232\\
 \qwCodeSeven& 0.8889& 0.9106& 0.9679& 0.9384& 0.9119&1.1739\\
\qwThreeZ   & 0.8512 & 0.8774 & 0.9498 & 0.9122 & 0.8914 & 1.3213 \\
\qwThreeOne   & 0.8664 & 0.9003 & 0.9457 & 0.9225 & 0.8964 & 0.6153 \\
\qwThreeFour     & 0.8807 & 0.9166 & 0.9471 & 0.9316 & 0.9073 & 0.7134 \\
\qwThreeEight     & 0.8843 & 0.9158 & 0.9533 & 0.9342 & 0.9092 & 0.7060 \\
\midrule
\textbf{AVG.} $\Delta$ & \textbf{0.8853} & \textbf{0.9109} & \textbf{0.9606} & \textbf{0.9351} & \textbf{0.9116} & \textbf{0.9969} \\
\bottomrule
\end{tabular}
\end{table}

%% file: tables/threshold_comparison.tex
\begin{table}[htbp]
  \centering
  \caption{Comparison of Learned vs. Fixed Entropy Thresholds on HumanEval+ (Pass@1). Fixed threshold is set to 1.2.}
  
  \renewcommand{\arraystretch}{1.25} 
  \setlength{\tabcolsep}{2pt}
    
  \resizebox{\linewidth}{!}{
    \begin{tabular}{l|ccccccccc|c}
      \toprule
       \textbf{Strategy} & \textbf{\dsOne} & \textbf{\dsSix} & \textbf{\stThree} & \textbf{\qwCodeOne} & \textbf{\qwCodeSeven} & \textbf{\qwThreeZ} & \textbf{\qwThreeOne} & \textbf{\qwThreeFour} & \textbf{\qwThreeEight} & \textbf{AVG} \\
      \midrule
      Fixed & 63.41\% & 73.17\% & 50.00\% & 65.85\% & 85.98\% & 21.34\% & 41.46\% & 51.22\% & 61.59\% & 57.11\%\\
      Learned & 65.24\% & 73.78\% & 50.61\% & 67.07\% & 87.20\% & 22.56\% & 44.51\% & 56.10\% & 62.80\% & 58.87\%\\
      \midrule
      $\Delta$ & +1.83\% & +0.61\% & +0.61\% & +1.22\% & +1.22\% & +1.22\% & +3.05\% & +4.88\% & +1.21\% & \textbf{+1.76\%} \\
      \bottomrule
    \end{tabular}
  }
  \label{tab:threshold_comparison}
\end{table}

%% file: figures/L_analysis.tex
\begin{figure}[htbp]
    \centering
    \includegraphics[width=0.7\columnwidth]{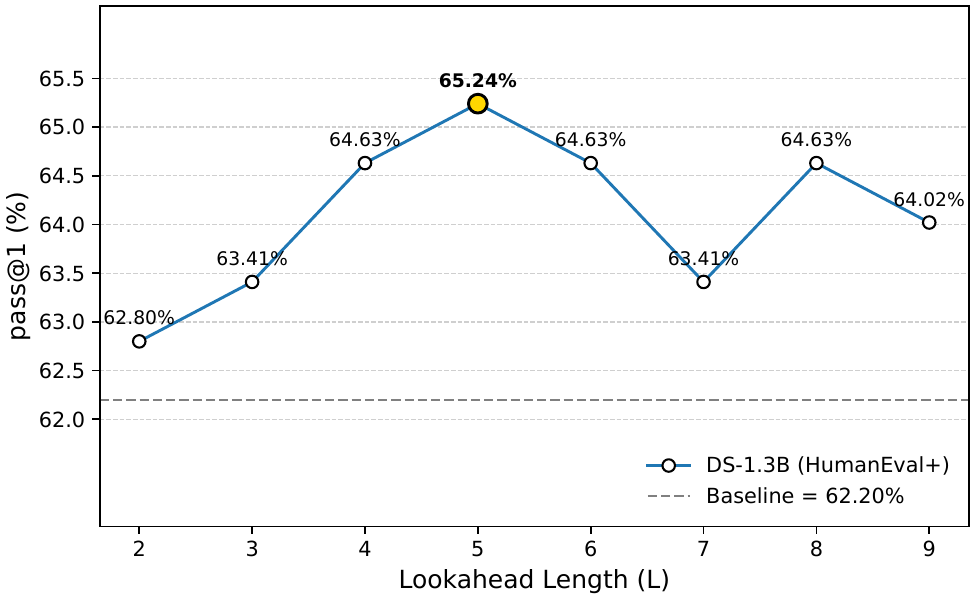}
    \Description{}
    \caption{Pass@1 Performance of the \dsOne Model on the HumanEval+ Dataset Under Different Lookahead Length (L) Configurations.}
    \label{fig:L_analysis}
\end{figure}

%% file: tables/sensitivity.tex
\begin{table}[htbp]
\centering
\caption{Sensitivity of \app performance (Pass@1) to entropy threshold ($\tau$) variations on HumanEval+. Evaluated with \dsOne. \textbf{Learned} denotes the optimal threshold derived from our approach.}
\label{tab:sensitivity}
\resizebox{0.70\columnwidth}{!}{
\begin{tabular}{lccc}
\toprule
Setting & Threshold $\tau$ & Pass@1 (\%) & Pause Rate (\%) \\
\midrule
Baseline (Greedy) & -- & 62.20 & -- \\
\arrayrulecolor{black!80}\midrule\arrayrulecolor{black}
$\tau^{LM} - 0.50$ & 0.4850 & 63.41 & 7.82 \\
$\tau^{LM} - 0.10$ & 0.8850 & 65.24 & 3.46 \\
$\tau^{LM} - 0.05$ & 0.9350 & 64.63 & 3.10 \\
\textbf{$\tau^{LM}$ (Learned)} & \textbf{0.9850} & \textbf{65.24} & \textbf{2.81} \\
$\tau^{LM} + 0.05$ & 1.0350 & 65.24 & 2.52 \\
$\tau^{LM} + 0.10$ & 1.0850 & 64.63 & 2.16 \\
$\tau^{LM} + 0.50$ & 1.4850 & 62.80 & 0.83 \\
\bottomrule
\end{tabular}
}
\end{table}

%% file: sections/discussion.tex
\section{Discussion}\label{sec:discussion}

\subsection{Generalizability and Transferability of the Learned Threshold (\texorpdfstring{$\tau^{LM}$}{tau-LM})}
\label{sec:discussion_threshold}

We conduct a quantitative analysis using \dsOne to evaluate the robustness of the learned entropy threshold ($\tau^{LM}$) across domain and prompt shifts, and provide guidance on its practical application.

\textbf{Robustness to Domain and Prompt Shifts.} 
We evaluate the transferability of $\tau^{LM}$ to library-heavy domains and varying prompt templates. We compute the optimal thresholds for two specific data science scenarios: a data science subset filtered from BigCodeBench (666 problems) and the DS-1000 benchmark~\cite{DS-1000} (1000 problems, featuring drastically different prompt styles and task requirements). The optimal thresholds for these two datasets are $0.9740$ ($\Delta = 0.0110$) and $0.9178$ ($\Delta = 0.0672$), respectively. Since our parameter sensitivity analysis (Section~\ref{sec:evaluation/RQ4}) demonstrates that deviations within $\pm 0.1$ do not significantly impact performance, a globally learned $\tau^{LM}$ remains highly effective without the need for task-specific tuning.

\textbf{Guidance on Relearning $\tau^{LM}$.} 
These findings indicate that $\tau^{LM}$ reflects the intrinsic uncertainty calibration of the base LLM rather than task-specific artifacts. Therefore, relearning $\tau^{LM}$ is generally unnecessary for different standard coding domains or prompt styles. Retraining is only recommended when the base model's confidence distribution changes fundamentally, such as after Supervised Fine-Tuning (SFT) or alignment techniques.

\subsection{Limitations}

While \app demonstrates improvements in code generation accuracy and maintains acceptable computational cost, it has several limitations that warrant further investigation:

\textbf{Reliance Solely on Shannon Entropy.}  
Our uncertainty trigger is based solely on the Shannon entropy of the token distribution, which provides a coarse measure of model uncertainty. More nuanced contextual features—such as entropy of subpopulations of tokens, token-level gradient norms, or higher-order statistics—might better capture subtle divergences. Future work could integrate richer uncertainty signals or leverage learned uncertainty estimators.

\textbf{Threshold Learning Simplicity.}  
We employ a logistic regression classifier to learn entropy thresholds, which offers interpretability but may be insufficiently expressive for capturing complex interactions among uncertainty features. More powerful models (e.g., small neural networks) could potentially improve threshold accuracy, at the cost of increased complexity.

\textbf{Fixed Lookahead Depth.}
Currently, \app utilizes a fixed lookahead length (e.g., $L=5$) across all benchmarks. While this serves as a reasonable global setting, repository-level tasks (e.g., DevEval) often require longer-range reasoning, and a short lookahead window may under-express these dependencies. Future research could explore adaptive, context-aware lookahead depths that dynamically adjust to accommodate both short-term syntactic constraints and long-term semantic reasoning demands.

Addressing these limitations will be an actionable direction for future work, towards making uncertainty-guided adaptive decoding more robust, flexible, and broadly applicable.

\subsection{Case Study}

To illustrate \app's practical advantage, Figure~\ref{fig:example} presents the \texttt{check\_dict\_case} task from HumanEval+. While standard greedy decoding unsafely assumes all dictionary keys are strings (causing runtime errors for integer keys), \app detects high entropy at this critical decision point. Consequently, it triggers lookahead reranking and selects a robust continuation that explicitly verifies key types (e.g., \texttt{isinstance(key, str)}) before case checking. This demonstrates how \app leverages entropy signals to proactively prevent subtle bugs and enhance generation reliability.

\input{figures/example}

%% file: figures/example.tex
\begin{figure}[htbp]
    \centering
    \includegraphics[width=0.7\columnwidth]{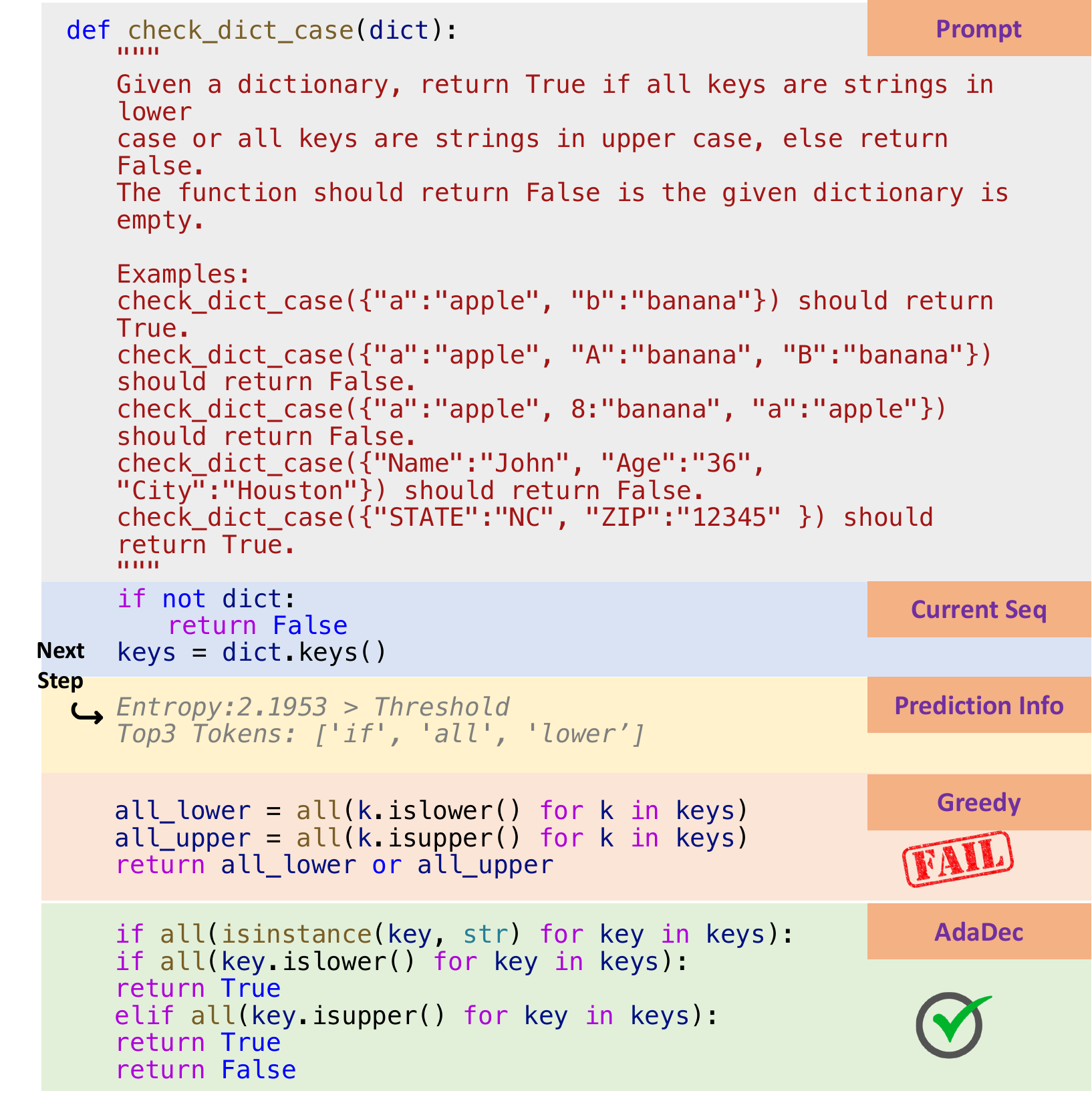}
    \Description{}
    \caption{A case study from HumanEval+}
    \label{fig:example}
\end{figure}

%% file: sections/threats.tex
\section{Threats to Validity}
We analyze threats to validity across four dimensions—internal, external, construct, and conclusion—and note our mitigation strategies or key limitations.

\textbf{Internal Validity.}  
Drift point identification may be imperfect: our semi-automated AST comparison could miss subtle semantic shifts (e.g., identifier drift), and human verification introduces potential annotator bias. We mitigated these risks through clear guidelines and cross-checking. Entropy-based analysis may also depend on tokenizer or model-specific distributions; to reduce this risk, we validated results across multiple models and confirmed entropy–rank correlations in RQ2.

\textbf{External Validity.}  
We focus on open-source models (DeepSeek, StableCode, Qwen) and three benchmarks (HumanEval+, MBPP+, DevEval). Results may not fully generalize to proprietary models or to broader real-world tasks. Future work should extend evaluation to more diverse datasets and model families.

\textbf{Construct Validity.}  
We evaluate functional correctness (Pass@1) and efficiency (latency), which directly measure the primary objectives of correctness and performance. Nevertheless, these metrics do not capture other important qualities such as readability or adherence to coding conventions. Similarly, our use of Shannon entropy as an uncertainty signal assumes it faithfully reflects model uncertainty, though it may only provide a coarse approximation. Alternative measures could offer complementary insights.

\textbf{Conclusion Validity.}  
Conclusion validity threats arise from the statistical robustness of our empirical results. Although we report consistent trends across multiple models and datasets, the sample size of benchmarks such as HumanEval+ remains relatively small, which may limit the statistical power of our findings. Additionally, hyperparameter choices (e.g., lookahead length $L$, candidate pool size $B$) may affect performance outcomes, and while we explored reasonable ranges, further sensitivity analysis would strengthen confidence in our conclusions.

%% file: sections/related_work.tex
\section{Related Work}

\subsection{LLM-based Code Generation}

Large Language Models (LLMs) have shown strong potential in software engineering tasks such as code generation~\cite{LiJia2023SKCODER,di2025enhancingcodegenerationbidirectional,tian2024fixinglargelanguagemodels,jiang2025rocodeintegratingbacktrackingmechanism,lin2024soen101codegenerationemulating}, unit test generation~\cite{Nan2025TestIntention,yang2024evaluationlargelanguagemodels,YuanEvaluat10.1145/3660783,ASystemLops2025,RyanCode-Aware2024}, and comment generation~\cite{haider2024promptingfinetuninglargelanguage,katzy2025qualitativeinvestigationllmgeneratedmultilingual,Lu2024ImprovingRetrieval}. Typically, LLMs generate code from textual requirements via prompt-based interaction. By learning expressive representations from large-scale code and documentation corpora, they can produce syntactically correct and context-aware code snippets~\cite{yuan2023evaluating, sun2024enhancing, ugare2024improving}. Open-foundation models further advance tasks like code completion, bug fixing~\cite{fu_vulrepair_2023, jin_inferfix_2023, xia_less_2022}, and test-case generation.

To evaluate their effectiveness, benchmarks like HumanEval~\cite{Humaneval}, MBPP~\cite{MBPP}, and DS-1000~\cite{DS-1000} focus on function-level generation, while ClassEval~\cite{Classeval} extends evaluation to class-level tasks~\cite{cao2026rigorreliabilityreproducibilitymatter}. More recently, CoderEval~\cite{Codereval} and DevEval~\cite{Deveval} target repository-level scenarios, reflecting growing research interest in applying LLMs to real-world software development.

\subsection{Decoding strategies for LLMs}

Large Language Models (LLMs) generate text in an autoregressive manner, producing one token at a time. At each generation step, the model samples a token from its predicted probability distribution. The most straightforward sampling strategy is greedy decoding, which selects the token with the highest probability as the next word. A more refined variant is beam search~\cite{beamsearch}, which maintains multiple candidate sequences at each decoding step and preserves the top-k sequences based on cumulative log-probabilities.

Beyond these, commonly used sampling strategies include temperature sampling~\cite{DBLP:conf/iclr/nsample}, top-p (nucleus) sampling~\cite{DBLP:conf/iclr/nsample}, and top-k sampling~\cite{DBLP:conf/iclr/nsample}, each designed to balance diversity and coherence in generated outputs.

To further improve sampling for code generation tasks, Li et al. introduced AdapT Sampling~\cite{AdapT} (Adaptive Temperature
Sampling). This method dynamically adjusts the temperature parameter T based on token context. For instance, when the current token initiates a code block, a higher T is used to encourage diversity and creativity. In contrast, for other tokens, a lower T is applied to suppress random noise and preserve syntactic and semantic correctness.

Another relevant approach is UnCert-CoT~\cite{UnCert-CoT}, which enhances code generation through uncertainty-aware Chain-of-Thought (CoT) reasoning, activating multi-path reasoning when token uncertainty surpasses a fixed threshold. While effective, its relies on fixed thresholds and costly CoT reasoning. In contrast, \app learns model-specific uncertainty thresholds for stronger generalization, while its lightweight rerank mechanism cuts computational costs significantly without compromising benchmark performance.

%% file: sections/conclusion.tex
\section{Conclusion}
This paper presents \app, an uncertainty-guided adaptive decoding framework for large language model-based code generation. Motivated by an empirical study revealing that ranking errors at high-uncertainty decoding steps often lead to logic drift, \app introduces a novel pause-then-rerank mechanism that dynamically identifies and corrects uncertain token predictions using Shannon entropy. By learning model-specific entropy thresholds and employing a lookahead-based reranking strategy, \app achieves substantial improvements in accuracy while maintaining computational cost and efficiency within acceptable bounds. Experiments on HumanEval+, MBPP+, and DevEval benchmarks demonstrate that \app consistently outperforms all compared methods. These findings underscore the importance of token-level uncertainty modeling and open up new directions for adaptive decoding in code generation and beyond.

%% file: sections/data_availability.tex
\section{Data Availability}

To facilitate the replication study, we have released our data and code at : \url{https://github.com/SYSUSELab/AdaDec}.

%% file: sections/acknowledgments.tex
\begin{acks}
This work was supported by the National Natural Science Foundation of China (Grant No.~92582117, Grant No.~62402113), the General Program of the Natural Science Foundation of Guangdong Province, China (Grant No.~2025A1515011631), and GMCC-SYSU Joint Lab for Smart Applications.
\end{acks}